\documentclass{llncs}

\usepackage[T1]{fontenc}
\usepackage[utf8]{inputenc}

\usepackage[british]{babel}
\usepackage{hyphenat}
\usepackage[usenames,dvipsnames]{color}  
\usepackage[pdfborder={0 0 0},colorlinks=true,urlcolor=blue,linkcolor=blue,citecolor=blue,bookmarks=true]{hyperref}

\hypersetup{pdftitle={Bandicoot: A Templated C++ Library for GPU Linear Algebra}}
\hypersetup{pdfauthor={Ryan R. Curtin, Marcus Edel, Conrad Sanderson}}

\usepackage{newtxtext}  
\usepackage{newtxmath}  
\usepackage{inconsolata}   

\usepackage{microtype}
\DisableLigatures[f,i,j,F,I,J]{encoding = *, family = *}

\usepackage[absolute]{textpos}
\usepackage{graphicx}
\usepackage[shortlabels]{enumitem}
\usepackage{bm}
\usepackage{booktabs}      
\usepackage{tabularx}
\usepackage{wasysym}

\begin{document}

\titlerunning{~}
\authorrunning{~}
\pagestyle{empty}

\title{\scalebox{0.90}{Bandicoot: A Templated C++ Library for GPU Linear Algebra}}


\author
  {
  Ryan R. Curtin\textsuperscript{{\tiny~}$\dagger$},
  Marcus Edel\textsuperscript{{\tiny~}$\ddagger$},
  Conrad Sanderson\textsuperscript{{\tiny~}$\ast\diamond$}
  }

\institute
  {
  \textsuperscript{$\dagger$}{\tiny~}\textit{NumFOCUS, USA;}~
  \textsuperscript{$\ddagger$}{\tiny~}\textit{Collabora, Canada;}~
  \textsuperscript{$\ast$}{\tiny~}\textit{CSIRO, Australia;}~
  \textsuperscript{$\diamond$}{\tiny~}\textit{Griffith University, Australia}
  }

\maketitle

\begin{abstract}

We introduce the Bandicoot C++ library for linear algebra and scientific computing on GPUs,
overviewing its user interface and performance characteristics,
as well as the technical details of its internal design.
Bandicoot is the GPU-enabled counterpart to the well-known Armadillo C++ linear
algebra library,
aiming to allow users to take advantage of GPU-accelerated computation
for their existing codebases without significant changes.
Exploiting similar internal template meta-programming techniques that Armadillo uses,
Bandicoot is able to provide compile-time optimisation of mathematical expressions within user code,
leading to more efficient execution.
Empirical evaluations show that Bandicoot can provide significant speedups over Armadillo-based CPU-only computation.
Bandicoot is available at {\url{https://coot.sourceforge.io}}
and is distributed as open-source software under the permissive Apache 2.0 license.

\begin{textblock}{9.1}(3.5,14.0)
\hrule
\vspace{1ex}
\noindent
\scalebox{0.736}{\textbf{Published in:} Lecture Notes in Computer Science (LNCS), Vol.{\tiny~}16465, pp.{\tiny~}156-168, 2026. DOI:{\tiny~}\href{https://doi.org/10.1007/978-981-95-9846-5_14}{\tt 10.1007/978-981-95-9846-5\_14}}
\end{textblock}

\end{abstract}

\section{Introduction}
\label{sec:introduction}

A number of recent significant advancements in scientific computing have been
enabled by the use of graphics accelerators / graphics processing units (GPUs),
notably in the fields of machine learning~\cite{ciresan2011flexible,Krizhevsky_2017},
astrophysics~\cite{hamada200942},
weather simulations~\cite{Bolt_2023,harris2003simulation},
data visualisation~\cite{stone2013gpu},
as well as healthcare and life sciences~\cite{ravi2016deep,vamathevan2019applications}.
GPUs are built on the concept of \textit{single-instruction-multiple-threads} (SIMT)
and are thus capable of performing the same operation on separate chunks of data in parallel.
Often, modern GPUs are extremely parallel:
Nvidia's recent mainstream consumer-focused GeForce RTX~5090 GPU contains 21,760 cores,
leading to a throughput of over 100 TFLOPS for 32-bit floating point data.

A large part of the reasons for the significant speedups and advancements in the fields listed above
is the core accelerations that GPUs provide to linear algebra operations.
The nature of most linear algebra computations is very well-suited to SIMT-like approaches:
for example, the common {\it axpy} vector-scalar product operation,
(ie., {\small $\mathbf{y} \gets \alpha \mathbf{x} + \mathbf{y}$})
is embarrassingly parallel and clearly suited for SIMT processing on a GPU.

However, from a user perspective,
writing scientific code that properly exploits GPUs is cumbersome and not portable.
The CUDA framework~\cite{nickolls2008scalable,nvidia_cuda} is specific to Nvidia devices,
and thus any CUDA code cannot directly work on competing AMD or Intel GPUs,
or any other GPU-like devices such as Google's tensor processing units~\cite{jouppi2017datacenter}.
Even higher-level languages such as MATLAB~\cite{Higham_2017} and R~\cite{Rmanual}
typically need special setup to use GPUs,
and that setup is typically limited to a certain type of GPU.
Standardisation efforts like OpenCL~\cite{stone2010opencl},
which promise code that works on any GPU-like device,
are often not well-supported by vendors.
Overall, balkanisation exists in the GPU computing space,
with numerous incompatible frameworks like
AMD's ROCm~\cite{amd_rocm} and
Apple's Metal~\cite{apple_metal},
in addition to Nvidia's CUDA.

From a non-expert user's perspective,
this situation is abysmal:
scientists and developers operating at a higher level do not want to spend time digging into low-level implementation details of their GPUs,
but instead want to be able to simply write linear algebra and have it work efficiently on their GPU.
This is, after all, the reason that linear algebra libraries and interfaces like BLAS~\cite{blackford2002updated} and LAPACK~\cite{anderson1999lapack} were originally written:
to provide an interface in order to abstract away the implementation details.

High-level interfaces have seen much success over the years,
specifically as they allow developers and scientists to maximise productivity by focusing on their higher-level task.
In this vein, frameworks like MATLAB and Mathematica~\cite{wolfram1999mathematica} have been extremely successful,
in large part because of the simple and intuitive interface they provide.
In the field of machine learning,
Python and its NumPy linear algebra add-on~\cite{harris2020array}
have become dominant in large part for the same reason.
However, these higher-level languages often come with various downsides;
for instance, MATLAB and Python have high amounts of overhead,
and require large runtime environments that depend on numerous additional packages.

Motivated by high-level MATLAB-style expressive syntax,
the well-known \mbox{Armadillo} C++ linear algebra library was created
to satisfy the need for a high-level and intuitive linear algebra interface
while maximising runtime performance and minimising overheads~\cite{sanderson2016armadillo,Sanderson_2025}.
Armadillo extensively uses template meta-programming~\cite{Vandevoorde_2017} in order to optimise linear algebra expressions at program compilation time;
this can provide significant and demonstrable speedups over other solutions~\cite{Psarras_2022}.
Its clear interface and flexibility has led to widespread adoption,
including via bindings to other languages like R~\cite{eddelbuettel2014rcpparmadillo} and Python~\cite{Rumengan_2021,Urlus_CARMA}.

However, Armadillo is only able to use CPUs and CPU-specific libraries
such as OpenBLAS~\cite{OpenBLAS} and Intel's Math Kernel Library (MKL) for its linear algebra operations.
Therefore, practitioners cannot make use of the GPUs that they likely have available to accelerate their linear algebra programs.
This motivated us to develop the \textit{Bandicoot} C++ linear algebra library for GPUs.

Bandicoot aims to provide an application programming interface (API) that closely matches Armadillo's API,
with the primary difference being that linear algebra operations are performed on the GPU instead of the CPU,
and the memory used for matrices is GPU memory instead of the host's system memory.
Bandicoot abstracts away the vendor-specific technologies that are used for GPU programming,
and can run on virtually any GPU via its support for both CUDA and OpenCL backends.
Users do not need to learn low-level details about their devices,
as Bandicoot will automatically configure and use the most appropriate device.

We continue the paper as follows.
An overview of Bandicoot's functionality is given in Section~\ref{sec:overview}.
We briefly describe Bandicoot's internal design in Section~\ref{sec:internals}.
Fundamental differences between Armadillo and Bandicoot are detailed in Section~\ref{sec:arma_comparison}.
Empirical speed comparisons demonstrating speedups gained by using Bandicoot are given in Section~\ref{sec:speed}.
A roadmap for future development of Bandicoot is listed in Section~\ref{sec:roadmap}.
An outlook and concluding statements are given in Section~\ref{sec:conclusion}.

\newpage
\section{Overview of Functionality}
\label{sec:overview}
\vspace{-1ex}

Bandicoot aims to provide an API that closely matches the API of the Armadillo C++ linear algebra library~\cite{sanderson2016armadillo}.
This means that Bandicoot's interface is the same as Armadillo's intuitive and readable interface,
so far as possible (see Section~\ref{sec:arma_comparison}).
An example Bandicoot program shown in Figure~\ref{fig:exampleprog} may therefore look familiar to Armadillo users.
In fact, due to Bandicoot's API-compatibility,
the code can be easily changed to be valid Armadillo code
by simply replacing {\small\tt \#include~<bandicoot>} with {\small\tt \#include~<armadillo>},
and {\small\tt using namespace coot} with {\small\tt using namespace arma}.
Conversely, many programs originally written to use Armadillo can be adapted to use Bandicoot with minimal changes.

\begin{figure}[!tb]
\centering
\fontsize{8.0pt}{9.0pt}\selectfont
\hrule
\vspace{1ex}
\begin{verbatim}
01: #include <bandicoot>
02: using namespace coot;
03: 
04: int main()
05:   {
06:   fvec A(1000, fill::randu);  // column vector with 32-bit floating point elements
07:   fvec B(1000, fill::randu);
08:   
09:   float result = sum(A);      // sum all elements of A
10:   
11:   B += 3 * A;                 // axpy operation
12:   
13:   fmat X(2000, 1000, fill::randu);   // matrix with 32-bit floating point elements
14:   fmat Y(2000, 1000, fill::randu);
15:   
16:   fmat Z = trans(X) * Y;  // matrix multiplication involving matrix transpose of X
17:   
18:   Z.diag() += 100;        // add scalar value to all elements on main diagonal
19: 
20:   fmat L, U, P;
21:   lu(L, U, P, Z);         // LU decomposition
22:   
23:   fvec C = solve(Z, A);   // solve system of linear equations
24:   
25:   fmat Xinv = pinv(X);    // Moore-Penrose pseudo-inverse based on SVD
26:   fmat Zinv =  inv(Z);    // standard inverse
27:   
28:   X.print("X:");          // pretty print matrix to std::cout stream
29:   
30:   return 0;
31:   }
\end{verbatim}
\vspace{-2ex}
\hrule
\vspace{-1ex}
\caption
  {
  Example C++ program using Bandicoot to perform linear algebra operations on~a~GPU.
  The corresponding Armadillo-based program that uses the~CPU instead of the~GPU
  can be obtained by replacing the code on line 1 with \mbox{\texttt{\scriptsize\#include <armadillo>}}
  and on line 2 with \mbox{\texttt{\scriptsize using namespace arma}}.
  }
\label{fig:exampleprog}
\end{figure}

The basic type in Bandicoot is the templated matrix class named \mbox{\small\tt Mat<element\_type>},
which as a template parameter takes the element type to be held in the matrix.
For example, the {\small\tt Mat<float>} class is a matrix with the {\small\tt float} element type that represents 32-bit floating point values.
Bandicoot provides convenient short-hand typedefs (aliases) for various matrix types;
thus, the {\small\tt fmat} class is equivalent to {\small\tt Mat<float>},
and {\small\tt dmat} class is equivalent to {\small\tt Mat<double>}.
Similar to the {\small\tt Mat} class,
Bandicoot provides the {\small\tt Col} and {\small\tt Row} classes for row and column vectors, respectively.
Corresponding convenience typedefs are provided,
such as {\small\tt fvec} and {\small\tt fcolvec} for {\small\tt Col<float>},
and {\small\tt frowvec} for {\small\tt Row<float>}.
Bandicoot also provides the {\small\tt Cube} class,
which acts as a ``3-dimensional matrix'' (quasi 3rd order tensor)
that is useful for higher-order operations that operate on collections of matrices.

When Bandicoot code is run,
linear algebra operations are performed on the GPU.
Bandicoot supports both CUDA and OpenCL backends.
Details of which backend to be used and which device to use
can optionally be controlled via the {\small\tt coot\_init()} function (shown in Figure~\ref{fig:coot_init}),
meant to be used at the beginning of a program.

\begin{figure}[!tb]
\centering
\fontsize{8.0pt}{9.0pt}\selectfont
\hrule
\vspace{1ex}
\begin{verbatim}
// Use only one of the following in your program, or omit entirely for automatic selection
coot_init("opencl", true /* print information */);
coot_init("opencl", false, 0 /* OpenCL platform ID */, 1 /* OpenCL device id */);
coot_init("cuda", true, 2 /* CUDA device ID */);
\end{verbatim}
\vspace{-2ex}
\hrule
\vspace{-1ex}
\caption
  {
  Example invocations of the {\tt coot\_init()} initialisation function to optionally select which backend and GPU device to use for Bandicoot linear algebra operations.
  }
\label{fig:coot_init}
\end{figure}

At the time of writing, Bandicoot provides over 100 functions and is comprised of about 85,000 lines of templated C++ code.
In addition to elementary arithmetic operations such as matrix addition and multiplication,
there are functions for statistics (such as mean and variance), 
signal processing (such as 1D and 2D convolution),
diagonal and submatrix views (allowing operations to use and be applied to parts of matrices),
broadcasting operations,
various matrix factorisations and decompositions
(including Cholesky decomposition, eigen decomposition, lower-upper (LU) decomposition, singular value decomposition),
matrix inverses (including pseudo-inverse),
and a solver for system of linear equations.
Wherever possible, the provided functions aim to be as similar as possible to MATLAB,
closely following the approach used by Armadillo~\cite{Sanderson_2025}.
A~comprehensive documentation of all available functionality in Bandicoot is provided online at \mbox{\small \url{https://coot.sourceforge.io/docs.html}}.

Bandicoot uses external support libraries whenever possible.
The employed libraries depend on which backend is enabled.
The {CUDA} backend uses cuBLAS, cuRand, and cuSolver and other parts of the CUDA Toolkit~\cite{nvidia_cuda}.
The {OpenCL} backend uses clBLAS~\cite{clblas} and internal adaptations of MAGMA~\cite{tdb10} and clMAGMA~\cite{Cao_2014},
which are essentially GPU versions of LAPACK~\cite{anderson1999lapack}.
This is similar to Armadillo's use of external support libraries such as OpenBLAS~\cite{OpenBLAS}, Intel MKL~\cite{Intel_MKL},
and other libraries that provide APIs compatible with BLAS and LAPACK.

Like Armadillo, Bandicoot provides a {\it wrapper library} to ease the linking process.
Linking to GPU libraries (especially in the CUDA framework) can be cumbersome,
with Nvidia even going so far as providing the {\it nvcc} compiler wrapper to simplify the process. 
Bandicoot's library obviates the need for this complexity,
and a typical program using Bandicoot can be compiled with a command as simple as:

{\footnotesize
\begin{verbatim}
    g++ prog.cpp -o prog -O2 -lbandicoot
\end{verbatim}
}

\noindent
The above command line links against the Bandicoot wrapper library
({\it libbandicoot.so} on Linux and {\it libbandicoot.dylib} on macOS),
which is dynamically linked against all of Bandicoot's dependencies.
In other words, the dependencies of the backends are abstracted away from the user,
simplifying development and deployment.

\section{Internal Design}
\label{sec:internals}
\vspace{-1ex}

Internally, Bandicoot is built on a template meta-programming framework similar to Armadillo's framework~\cite{Sanderson_2025}.
The underlying idea is to provide significant compile-time and run-time optimisations of mathematical expressions present in user code.
This is accomplished by automatically collecting (at compile time)
the structure of a linear algebra expression as an elaborate custom type,
and then, wherever possible,
using techniques such as template specialisation~\cite{Vandevoorde_2017}
to execute the expression in a more efficient manner.
Importantly, to keep Bandicoot user-friendly,
this entire template meta-programming framework is internal to the library,
and therefore invisible to the user.

All matrix and vector objects as well as all linear algebra expressions inherit from the fundamental class {\small\tt Base<eT,T1>},
where {\small\tt eT} refers to the {\it element type} and {\tt T1} refers to a specific class representing a matrix, vector, or compound linear algebra expression.
This is a form of static polymorphism, omitting runtime virtual-table lookups~\cite{alexandrescu2001modern,Vandevoorde_2017}.
User-facing classes include the {\small\tt Mat<eT>}, {\small\tt Col<eT>}, and {\small\tt Row<eT>} classes,
which represent matrices, column vectors, and row vectors, respectively.
The vector classes are derived from the matrix class, which allows the treatment of vectors as specialised versions of matrices.
A number of predefined short-hand {\tt typedef}s are given,
allowing users to use more concise forms such as {\tt fmat} type instead of {\small\tt Mat<float>}.

Expressions are represented as C++ types using the following template classes:
\mbox{\small\tt Op<eT,T1,op\_type>},
\mbox{\small\tt eOp<eT,T1,eop\_type>},
\mbox{\small\tt Glue<eT,T1,T2,glue\_type>},
and
\mbox{\small\tt eGlue<eT,T1,T2,eglue\_type>}.
Each of these classes represents a specific type of linear algebra operation:

\begin{enumerate}[{$\bullet$},leftmargin=*]
  \item \mbox{\small\tt Op<T1,op\_type>}: an operation of type {\small\tt op\_type} that operates on an object with arbitrary Bandicoot type {\small\tt T1} that may result in a change of dimensions (eg., transposing a matrix) 
  \item \mbox{\small\tt eOp<T1,eop\_type>}: an element-wise operation of type {\small\tt eop\_type} on object with type {\small\tt T1} that preserves matrix dimensions (eg.,~multiplying matrix by scalar)
  \item \mbox{\small\tt Glue<T1,T2,glue\_type>}: an operation of type {\small\tt glue\_type} that operates on two objects with corresponding types {\small\tt T1} and {\small\tt T2} that may result in a change of dimensions (eg.,~matrix multiplication)
  \item \mbox{\small\tt eGlue<T1,T2,eglue\_type>}: an element-wise operation of type {\small\tt eglue\_type} on two objects with corresponding types {\small\tt T1} and {\small\tt T2} that preserves matrix dimensions (eg.,~adding two matrices with same dimensions)
\end{enumerate}

\noindent
The operation types (such as \mbox{\small\tt op\_type} and \mbox{\small\tt glue\_type})
are template types that represent the operation to be performed.
For example, the transpose operation is represented as \mbox{\small\tt op\_trans}, while the matrix multiplication operation is denoted as \mbox{\small\tt glue\_times}.
 
Representations of expressions via the above types is automatically performed as users call functions.
Importantly, Bandicoot's functions generally do not return matrix types such as \mbox{\small\tt Mat<eT>},
and instead return any of the expression types listed above,
which capture (as a type) the expression to be performed.
As an example, the function signature for Bandicoot's \mbox{\small\tt trans()} function (representing matrix transpose) is as follows:

{\scriptsize
\begin{verbatim}
template<typename eT, typename T1> inline const Op<T1, op_trans> trans(const Base<eT, T1>& expr);
\end{verbatim}
}

There are two important characteristics of the above function signature:
{\bf (i)} the function can receive any arbitrary Bandicoot expression and is not limited to matrix or vector objects as the input; 
{\bf (ii)} the function does not return a matrix or vector object,
and instead returns an \mbox{\small\tt Op<T1,op\_trans>} object which indicates that execution will require taking the transpose of {\small\tt T1}, where {\small\tt T1} is the given arbitrary Bandicoot expression.

Multiple linear algebra operations can be combined into a compound expression.
Through Bandicoot's overloading the {\tt *} operator function,
an expression involving matrix transpose and multiplication can be written by the user 
in a straightforward manner as
{\small \verb|trans(X) * Y|},
which is automatically internally represented by the following compound type:
{\small \verb|Glue< Op<Mat<float>, op_trans>, Mat<float>, glue_mul>|}.

Bandicoot expressions are only evaluated when the entire expression is assigned to matrix or vector object through the {\small\tt =} operator.
For example, {\small \verb|fmat Z = trans(X) * Y|}.
This is known as {\it delayed evaluation}
and is distinct from standard {\it eager evaluation}~\cite{Watt_2004}.
Delayed evaluation approaches have been shown to provide significant speedups over eager evaluation of expressions~\cite{Psarras_2022}.
Instead of evaluating each component of the expression separately, 
Bandicoot automatically analyses the entire expression 
and uses compile-time pattern matching (via template specialisations)
to choose the minimal set of calls to appropriate GPU functions or kernels that implement the expression.

For the case of the {\small \verb|trans(X) * Y|} expression,
Bandicoot maps the expression as a {\it single} call to the optimised {\small\tt GEMM} matrix multiplication function present in the cuBLAS (CUDA) and clBLAS (OpenCL) backends.
The {\small\tt GEMM} function takes transpose indicators as part of its long argument list,
meaning that the transpose operation is only done {\it implicitly}, which in turn reduces execution time.
Similar optimisations are performed for many other compound expressions, such as the {\it axpy} operation.

Each of the operation classes ({\small\tt Op}, {\small\tt Glue}, etc) is deliberately lightweight,
holding only constant references to its operands and possibly a few integers of metadata.
As such, modern C++ compilers are able to determine that instances of these classes are ephemeral and only need to exist at compile-time.
This allows the compilers to produce compact machine code where instances of these classes are essentially optimised away,
while retaining the generated sequence of calls to GPU functions and kernels.

\section{Fundamental Differences Between Armadillo and Bandicoot}
\label{sec:arma_comparison}
\vspace{-1ex}

Although Bandicoot aims to be API-compatible with Armadillo wherever possible,
major differences in the computational designs of CPUs and GPUs mean that
Bandicoot and Armadillo have some fundamental differences, as elucidated below.

The first fundamental difference is that consumer-grade GPUs typically tend to provide considerably higher performance for 32-bit (and lower) precision floating point
than for 64-bit floating point~\cite{haidar2018harnessing,whitehead2011precision}.
For this reason, when using Bandicoot,
it is generally recommended to use matrices and vectors with the {\it float} element type instead of the {\it double} element type.
That is, the use of {\tt fmat} matrix type is generally recommended instead of {\tt dmat} matrix type.

The second fundamental difference is related to the amount of parallelism that GPUs provide.
The main avenue to speedup with a GPU is provided by the extreme level of parallelism.
For instance, Nvidia's recent GeForce RTX~5090 GPU contains 21,760 cores,
whereas recent AMD Epyc (x86-64 compatible) CPUs contain up to 192 cores
(i.e.,~about 2~orders of magnitude difference).
The clear implication of this is that keeping all of the cores in the GPU busy
is of the utmost importance to attain faster operation.
For this reason, Bandicoot is ill-suited to calculations on small matrices.
In our empirical evaluations shown in Section~\ref{sec:speed},
Bandicoot is observed to be faster than the CPU for matrices sized roughly {\small $1000 \times 1000$} or larger when using 32-bit floating point elements.
It~must be noted that the matrix size crossover point depends on the nature and complexity of the computational workload,
and is also likely to vary depending on device specifics (eg., CPU type, GPU type, memory type, etc).

The third fundamental difference is that a large set of predefined GPU kernels must be compiled at runtime for each specific device.
With Armadillo, all operations can be compiled entirely at compile time.
In contrast, when any Bandicoot program is run,
the first step is to compile the entire set of predefined GPU kernels for the device that is found.
Bandicoot caches the compiled kernels,
so on most systems this startup step is a one-time cost that typically takes a few minutes.

The fourth major fundamental difference is the degree of expression optimisation.
Whereas Armadillo is able to induce the C++ compiler to generate CPU code highly tailored to each individual expression~\cite{Psarras_2022,Sanderson_2025},
Bandicoot has fewer opportunities to do so
as it (currently) must choose a sequence of calls from a set of predefined GPU functions and kernels.

The last difference is the location where matrix and vector data is stored.
In Armadillo, data is stored in host/system memory, while in Bandicoot the data is stored in device (GPU) memory.
Using Bandicoot in C++ programs means that the control code runs on the host,
which in turn means that accessing the individual elements of a matrix requires a transfer from GPU memory to host memory.
For a single element, this is computationally expensive and inefficient.
For this reason, C++ iterators over Bandicoot matrices are not provided.
However, for situations where mixed use of Armadillo and Bandicoot is appropriate,
the {\small\tt conv\_to()} function can be used for converting between matrices stored on the host and the GPU,
as shown in Figure~\ref{fig:convto}.

\begin{figure}[!tb]
\centering
\fontsize{8.0pt}{9.0pt}\selectfont
\hrule
\vspace{1ex}
\begin{verbatim}
coot::fmat X_gpu(1000, 1000, coot::fill::randu);

// transfer matrix from GPU to CPU
arma::fmat X_cpu = coot::conv_to<arma::fmat>::from(X_gpu);

// transfer back from CPU to GPU
X_gpu = coot::conv_to<coot::fmat>::from(X_cpu);
\end{verbatim}
\vspace{-2ex}
\hrule
\vspace{-1ex}
\caption
  {
  Code fragment demonstrating conversion between matrices stored in host/system memory and device (GPU) memory.
  The {\tt coot::} and {\tt arma::} prefixes invoke C++ namespaces,
  and are used here to disambiguate between Bandicoot and Armadillo matrices/functions.
  }
\label{fig:convto}
\end{figure}

\newpage
\clearpage

\begin{figure}[!tb]
  \centering
  \begin{minipage}{0.49\textwidth}
    \centering
    \includegraphics[width=\textwidth]{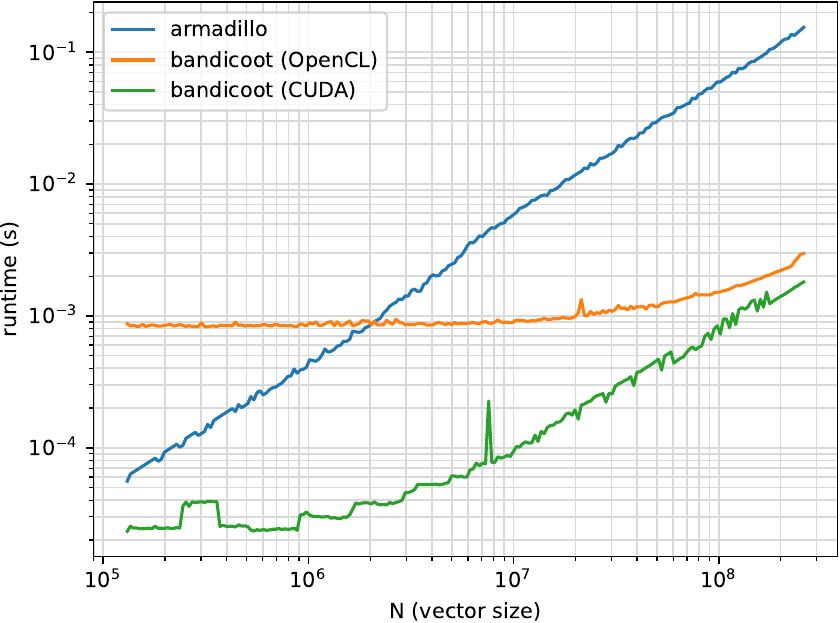}\\
    {\scriptsize {\tt ~~~~~}\textbf{(a)}}
  \end{minipage}
  \hfill
  \begin{minipage}{0.49\textwidth}
    \centering
    \includegraphics[width=\textwidth]{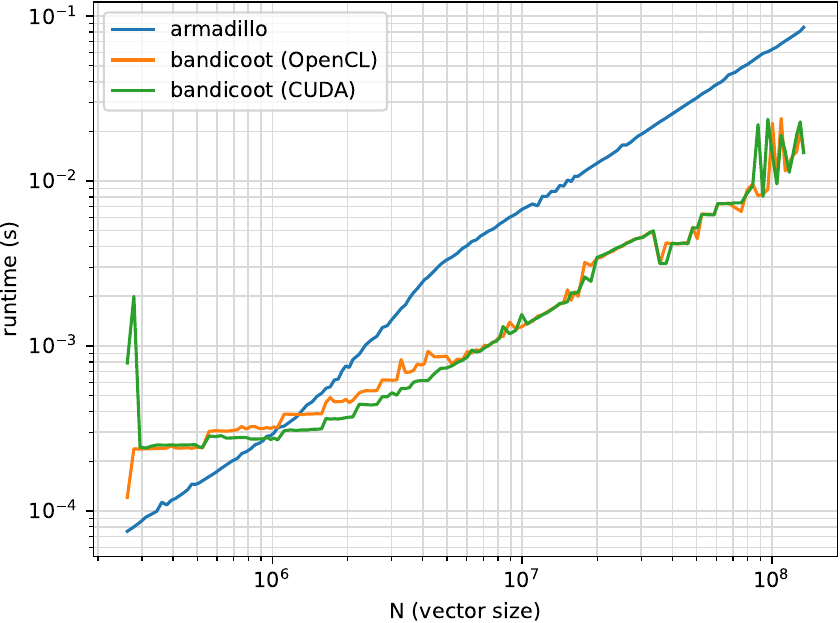}\\
    {\scriptsize {\tt ~~~~~}\textbf{(b)}}
  \end{minipage}
  \vspace{-2ex}
  \caption{Performance of Bandicoot and Armadillo as a function of vector size (N) on the tasks of: \textbf{(a)} summation, and \textbf{(b)} {\it axpy} operation. 32-bit floating point values are used.}
  \label{fig:vector_ops}
  \vspace{1ex}
\end{figure}

\begin{figure}[!tb]
  \centering
  \begin{minipage}{0.49\textwidth}
    \centering
    \includegraphics[width=\textwidth]{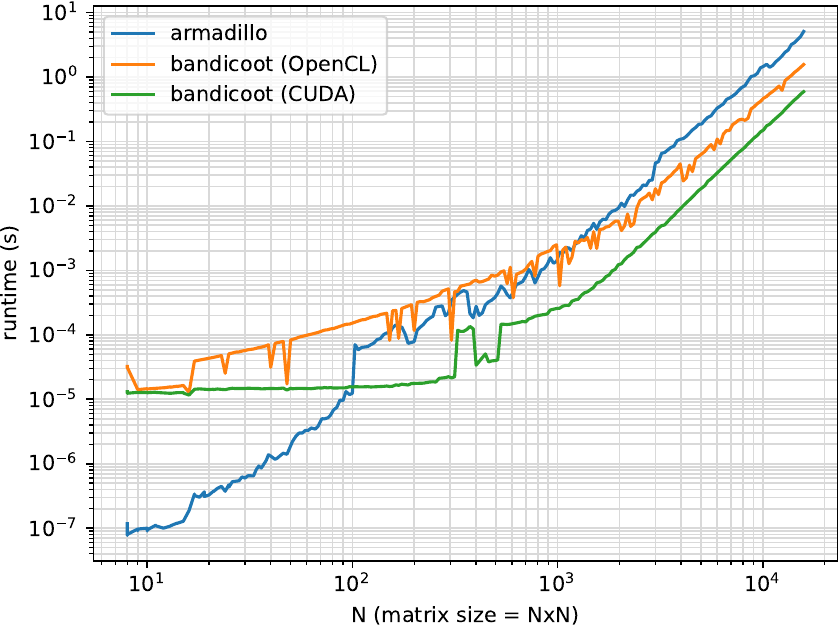}\\
    {\scriptsize {\tt ~~~~~}\textbf{(a)}}
  \end{minipage}
  \hfill
  \begin{minipage}{0.49\textwidth}
    \centering
    \includegraphics[width=\textwidth]{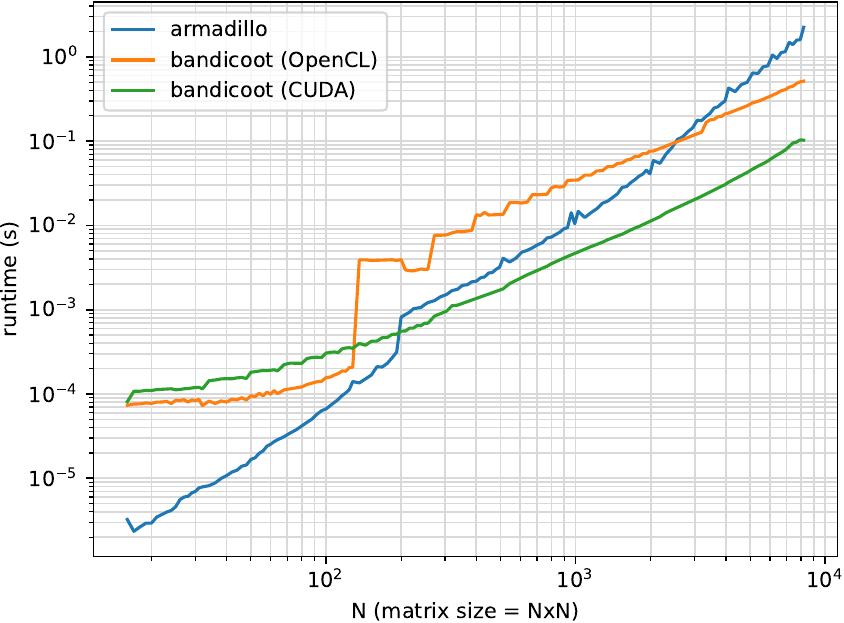}\\
    {\scriptsize {\tt ~~~~~}\textbf{(b)}}
  \end{minipage}
  \vspace{-2ex}
  \caption{Performance of Bandicoot and Armadillo as a function of matrix size (N{$\times$}N) on the tasks of: \textbf{(a)} matrix multiplication, and \textbf{(b)} LU decomposition. 32-bit floating point values are used.}
  \label{fig:matrix_ops}
\end{figure}

\section{Empirical Speed Comparison}
\label{sec:speed}
\vspace{-1ex}

We performed comparative speed evaluations in order to demonstrate the speedups that can be obtained when using Bandicoot.
The timing comparisons focus on the speedups Bandicoot can provide over Armadillo due to the use of GPUs instead of CPUs.
The evaluations were done on four tasks with progressively increasing computational complexity:
{\bf (1)}~summation of all elements in a vector,
{\bf (2)}~the {\it axpy} vector-scalar product operation ({\small $\mathbf{y} \gets \alpha \mathbf{x} + \mathbf{y}$}),
{\bf (3)}~matrix multiplication,
{\bf (4)}~LU~decomposition.
Example invocations of the above operations are given in Figure~\ref{fig:exampleprog}.

All comparisons were done on a system with an Intel Core i9-10920X
(3.5~GHz, 12~cores, 19.25MB~cache),
an Nvidia RTX~2080Ti GPU (1350 MHz, 4352 cores, 11GB RAM), 
and 128GB system RAM, with Linux kernel 6.1 and the GCC 12 C++ compiler.
Comparisons were performed using matrices 32-bit floating-point precision floating point values;
see Section~\ref{sec:arma_comparison} for more details on precision limitations in GPUs.
Armadillo used OpenBLAS~\cite{OpenBLAS} as the optimised implementation of BLAS and LAPACK.
Bandicoot was evaluated with the OpenCL and CUDA backends separately.

All results are reported in terms of seconds and exclude startup costs (eg., matrix generation and GPU kernel compilation).
Figure~\ref{fig:vector_ops} shows the results for the vector based operations (tasks 1 and 2),
using vector sizes from about {\small $10^{5}$} to about {\small $10^{8}$} elements.
Figure~\ref{fig:matrix_ops} shows the results for the matrix based operations (tasks 3 and 4),
using square matrix sizes from about {\small $10 \times 10$} to about {\small $10,000 \times 10,000$} elements.

For the straightforward summation operation (task 1),
Bandicoot with the OpenCL backend is faster than Armadillo for vector sizes larger than about {\small $2\times10^{6}$} elements.
Bandicoot with the CUDA backend is always faster than Armadillo,
which is not unexpected on an Nvidia device.
Under both OpenCL and CUDA backends, Bandicoot can outperform Armadillo by roughly two orders of magnitude for very large vector sizes.

For the {\it axpy} operation (task 2), Bandicoot with either the OpenCL or CUDA backend is faster than Armadillo
when the vector size is larger than about {\small $10^{6}$} elements.
Both OpenCL and CUDA backends yield similar performance on this operation,
and attain a speedup factor of around 10 for very large vector sizes.

For matrix multiplication (task 3), Bandicoot with the OpenCL backend is noticeably faster than Armadillo
once the matrix size is larger than the crossover point of about {\small $1000 \times 1000$} elements.
For Bandicoot with the CUDA backend, the crossover point is lower at about {\small $100 \times 100$} elements,
with performance about ten times faster than Armadillo for larger matrix sizes.

For the relatively computationally intensive LU decomposition (task 4),
Bandicoot with the OpenCL backend starts to become faster than Armadillo for matrices larger than about {\small $2500 \times 2500$} elements.
For Bandicoot with the CUDA backend, the corresponding crossover point is lower at about {\small $200 \times 200$} elements;
for matrix sizes around {\small $8000 \times 8000$}, the speedup is approximately a factor of 10.

\section{Development Roadmap}
\label{sec:roadmap}
\vspace{-1ex}

We are continuing to further develop Bandicoot, with the aims detailed below.

\begin{enumerate}[{$\bullet$},leftmargin=*]
\itemsep=0.5ex

\item
{\bf API expansion}.
The primary goal for Bandicoot is to be API-compatible with Armadillo as much as possible,
to ease adaptation of Armadillo based programs.
To that end, we are currently in the process of expanding Bandicoot's API with
additional {matrix decompositions},
such as QR, QZ, Schur and generalised eigenvalues~\cite{Golub_2013,Trefethen_2022},
as well as expansion of the {\tt solve()} function to match the multitude of options present in Armadillo.
Furthermore, we are adding support for {complex numbers}, to allow matrices such as {\small\tt Mat<~std::complex<float>~>}.

\item
{\bf Use of CLBlast instead of clBLAS}.
The clBLAS project~\cite{clblas} has not been updated in many years
and has bugs that are likely to remain unfixed.
The newer and better maintained \textit{CLBlast} project
provides an appealing higher-performance alternative implementation of BLAS functions on OpenCL~\cite{nugteren2018clblast}.
Like Armadillo's support for any BLAS/LAPACK replacement,
the goal for Bandicoot is to allow the use of either library with the OpenCL backend.

\newpage

\item
{\bf Additional backends}.  The balkanisation of the GPU space is a real
concern, as OpenCL is effectively being replaced by proprietary frameworks from multiple vendors.
Although OpenCL is likely to work as a backend for many years,
higher performance can be obtained with vendor-specific backends
due to the amount of effort that vendors put into optimising the backends for their hardware.
Examples of useful backends include:
AMD's ROCm~\cite{amd_rocm}, Apple's Metal~\cite{apple_metal}, and Intel-backed oneAPI~\cite{OneAPI}.

\item
{\bf Low-precision type support}.
Low-precision floating-point types are often used to increase computational
performance and/or energy efficiency~\cite{hubara2017quantized,tagliavini2018transprecision}.
Types such as \textit{brain floating point} have also gained widespread adoption~\cite{Burgess_2019,Henry_2019}.
CUDA supports precisions down to 8-bit ({\small\tt fp8}),
and OpenCL extensions support half-precision floating-point ({\small\tt fp16}).

\end{enumerate}

\section{Conclusion}
\label{sec:conclusion}
\vspace{-1ex}

We have provided an overview of the open-source Bandicoot C++ library for linear algebra on GPUs,
which aims to provide an API that is compatible with the CPU-focused Armadillo C++ linear algebra library.
This allows users to take advantage of GPU-accelerated computation for their existing codebases without significant changes,
considerably reducing development time and lowering the risk of introducing bugs.
Bandicoot is licensed under the permissive Apache 2.0 license~\cite{Li_2025,Laurent_2004}.

Empirical evaluations show that Bandicoot can provide significant speedups over CPU-only computation,
depending on the nature of the workload and the size of the data.
Bandicoot can use either an OpenCL or CUDA backend,
with plans to support further backends such as AMD's ROCm.

In follow-up work, we describe an extended template metaprogramming framework
to generate fused GPU kernels directly at compile-time,
providing considerable speedups at run-time~\cite{Curtin_2026b}.
Work is currently underway to adapt higher-level libraries relying on Armadillo
to be also capable of using Bandicoot and hence provide GPU-accelerated algorithms.
Two important examples are the
\textit{ensmallen} numerical optimisation library~\cite{curtin2021ensmallen}
and the
\textit{mlpack} machine learning library~\cite{curtin2023mlpack}.

Bandicoot's open-source nature enables extensions and customisation,
and we hope that the community will find it useful.
Contributions are welcome and can be made at the Bandicoot code repository:
\mbox{\small\url{https://gitlab.com/bandicoot-lib/bandicoot-code/}}.
More details about the library,
including comprehensive documentation of user-accessible functions and classes,
can be found on the publicly accessible website~at \mbox{\small\url{https://coot.sourceforge.io}}.

\section*{Acknowledgements}
\vspace{-1ex}

{\small
Ryan Curtin's contributions are based on work supported
by the National Aeronautics and Space Administration (NASA)
under the ROSES-23 HPOSS program, grant no.~80NSSC24K1524.
}

\newpage

\def~{\,}  

\bibliographystyle{splncs04}
\bibliography{refs}

\end{document}